\documentclass{article}
\usepackage{arxiv}
\usepackage[utf8]{inputenc} 
\usepackage[T1]{fontenc}    
\usepackage{url}            
\usepackage{booktabs}       
\usepackage{amsfonts}       
\usepackage{nicefrac}       
\usepackage{microtype}      
\usepackage{lipsum}
\usepackage{graphicx}
\usepackage{amsmath}

\title{Real-time Monitoring of Autonomous Vehicle's Time Gap Variations: A Bayesian Framework}

\author{
  Wissam Kontar \\
  Department of Civil and Environmental Engineering\\
  University of Wisconsin-Madison\\
  Madison, WI 53706 \\
  \texttt{kontar@wisc.edu} \\
   \And
 Soyoung Ahn \thanks{Corresponding author. \emph{This paper was accepted to the 99th Annual Meeting of the Transportation Research Board, Washington, D.C., United States, 2020 } } \\
  Department of Civil and Environmental Engineering\\
  University of Wisconsin-Madison\\
  Madison, WI 53706 \\
  \texttt{sue.ahn@wisc.edu} \\
}

\begin{document}
\maketitle

\begin{abstract}
This paper proposes a novel monitoring methodology for car-following control of automated vehicles that uses real-time measurements of spacing and velocity obtained through vehicle sensors. This study focuses on monitoring the time gap, a key parameter that dictates the desired following spacing of the controlled vehicle. The goal is to monitor deviations in actual time gap from a desired setting and detect when it deviates beyond a control limit. A random coefficient modeling is developed to systematically capture the stochastic distribution of the time gap and derive a closed-form Bayesian updating scheme for real-time inference. A control chart is then adopted to systematically set the control limits and inform when the time gap setting should be changed. Simulation experiments are performed to demonstrate the effectiveness of the proposes method for monitoring the time gap and alerting when the parameter setting needs to be changed.
\end{abstract}


\section{\uppercase{Introduction}}
Automated vehicles (AV) have become a technological focus in the pursuit of efficient and safe transportation. Currently, the technology spans across two major categories: connected and automated vehicles (CAVs) and automated vehicles. CAVs are complemented by the ability to communicate and share information, thus facilitating cooperative control. On the other hand, automated vehicles make decision based only on information from their sensors. These technologies are exemplified in the market through autonomous systems such as Adaptive Cruise Control (ACC) and Cooperative Adaptive Cruise Control (CACC). 

To harvest the full potential of AV technology for improving traffic stability and capacity, recent field studies have shown the direct implications of car-following control for traffic improvements \cite{milanes2013cooperative,perez2013cooperative,shladover2012impacts}. Specifically, through car-following control, shorter time gaps between vehicles can be realized while effectively regulating and resolving any disturbances (i.e. stop-and-go, aggressive deceleration/acceleration). Numerous studies have developed car-following control algorithms, which can be generally categorized intro: (i) linear state feedback/feedforward controllers \cite{swaroop1994comparision,swaroop1996string}; (ii) optimal control that facilitates online optimization of a pre-designed objective function over a future time horizon, incorporating current system measurements and dynamics (e.g., Model Predictive Control (MPC)) \cite{wang2014rolling,zhou2017rolling}; (ii) artificial intelligence (AI) base controller that adopts data-driven algorithms \cite{zhang2017consensus}. 

Perhaps the most extensively adopted controller in both ACC and CACC systems is the linear feedback/feedforward controller, which enjoys a high flexibility and simplicity in formulating the control strategy and incorporating uncertainty. Notably, the linear controller takes on a hierarchical form where an upper controller dictates the car-following policy that is then fed into the lower controller to prescribe the acceleration rate needed. The policy for the upper controller is either constant time headway (CTH) or constant spacing (CS) policy. CTH defines the equilibrium spacing as a linear relationship between spacing and velocity \cite{rajamani2001experimental}, while the CS policy dictates a constant time invariant spacing \cite{darbha1999intelligent}. Of the two, CTH is gaining more acceptance as it is robust against disturbance propagation and is consistent with the normal driving intuition (i.e., a driver is likely to slow down when spacing decreases) \cite{li2002traffic}. The Society of Automotive Engineers (SAE) now recommends the CTH policy as a common standard in the current ACC/CACC systems with linear controllers. 

The basic idea behind the CTH policy is to regulate the vehicle's longitudinal movement (acceleration/deceleration rate) as to maintain a desired spacing. Specifically, the desired spacing, at each control time instant, is function of a pre-defined constant time gap parameter setting, speed and a standstill spacing. For instance, in an optimal setting we expect the controlled vehicle to maintain perfectly the desired spacing at all times. However, empirical experiments suggest that high gap errors and fluctuations in car-following behavior result in poor tracking of time gap setting, thus leading to undesired control outputs \cite{milanes2014modeling}. Additionally, the value of time gap setting has been under the spotlight in current literature especially as it impacts string stability and disturbance propagation. 

In a series of field experiments performed on ACC and CACC systems by the California PATH program, variations in the actual time gap profile were noted as compared to the desired time gap setting \cite{shladover2009effects}. Specifically, when the leading vehicle undergoes repeated oscillations of stop-and-go, the ACC system experienced significant gap errors. In cases of car platooning, the actual time gap profile shows overshooting, which could result in driver discomfort and may lead to string instability. In contrast, CACC systems performed significantly better under disturbances due to its communication ability, yet variations in time gap are still present, possibly due to uncertainties in system dynamics and sensor measurements (e.g., air drag, communication delay, measurement noise). While researchers have studied and incorporated some uncertainty in formulating the control system, uncertainties specific to the time gap parameter remain unaddressed. The conjuncture here, is that uncertainties in time gap are reflective in the overall vehicle's performance and are hard to model without any real knowledge into their nature. Thus, the need to real-time monitoring systems. These allow us to gain more insights into the real-time performance of the AV and take decisions into its control parameter settings, for instance time gap. 

Based on the above insights, this paper proposes a novel direction in assessing the performance of the control systems, which stresses on the importance of coupling the vehicle control system with a monitoring system able to reason about its condition in real-time. Specifically, the goal of this paper is to develop a monitoring framework to examine the variations in time gap parameter informed from real-time sensor data. Accordingly, we introduce a random coefficient formulation of the physical car-following model with a Bayesian updating scheme. Such formulation enjoys high flexibility and analytical properties that allow us to capture the stochastic in time gap parameter. The control charts are introduced to determine the feasible region for time gap variations. Thus, the proposes monitoring system can inform when the time gap should be changed to attain more stable performance.

\section{Model Formulation}
\label{sec:2}
In this section we present the formulation for monitoring time gap parameter and derive a closed form of the Bayesian updating scheme, informed by real-time sensor data. Furthermore, we present the formulation of Shewart-univariate control charts. 

\subsection{Background}
\begin{figure}[!htb]
    \centering
    \centerline{\includegraphics[width=0.6\columnwidth]{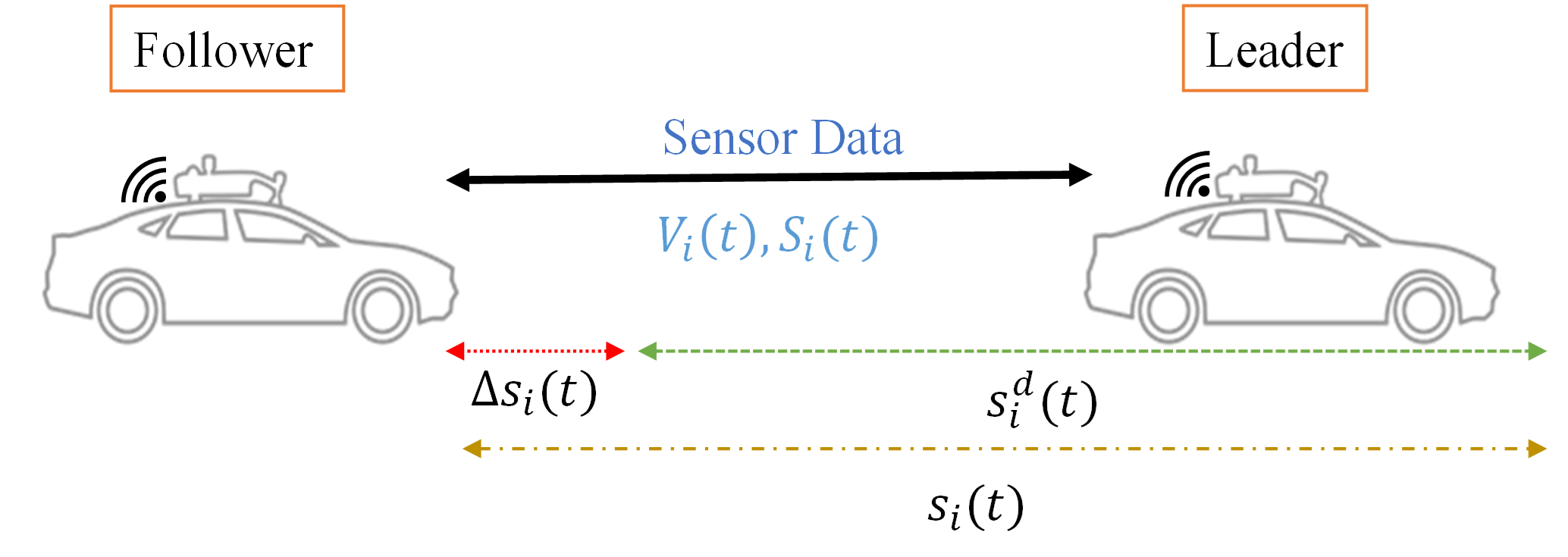}}
    \caption{Car-following control schematic}
    \label{fig:illustration}
\end{figure}

\begin{figure}[!htb]
    \centering
    \centerline{\includegraphics[width=0.6\columnwidth]{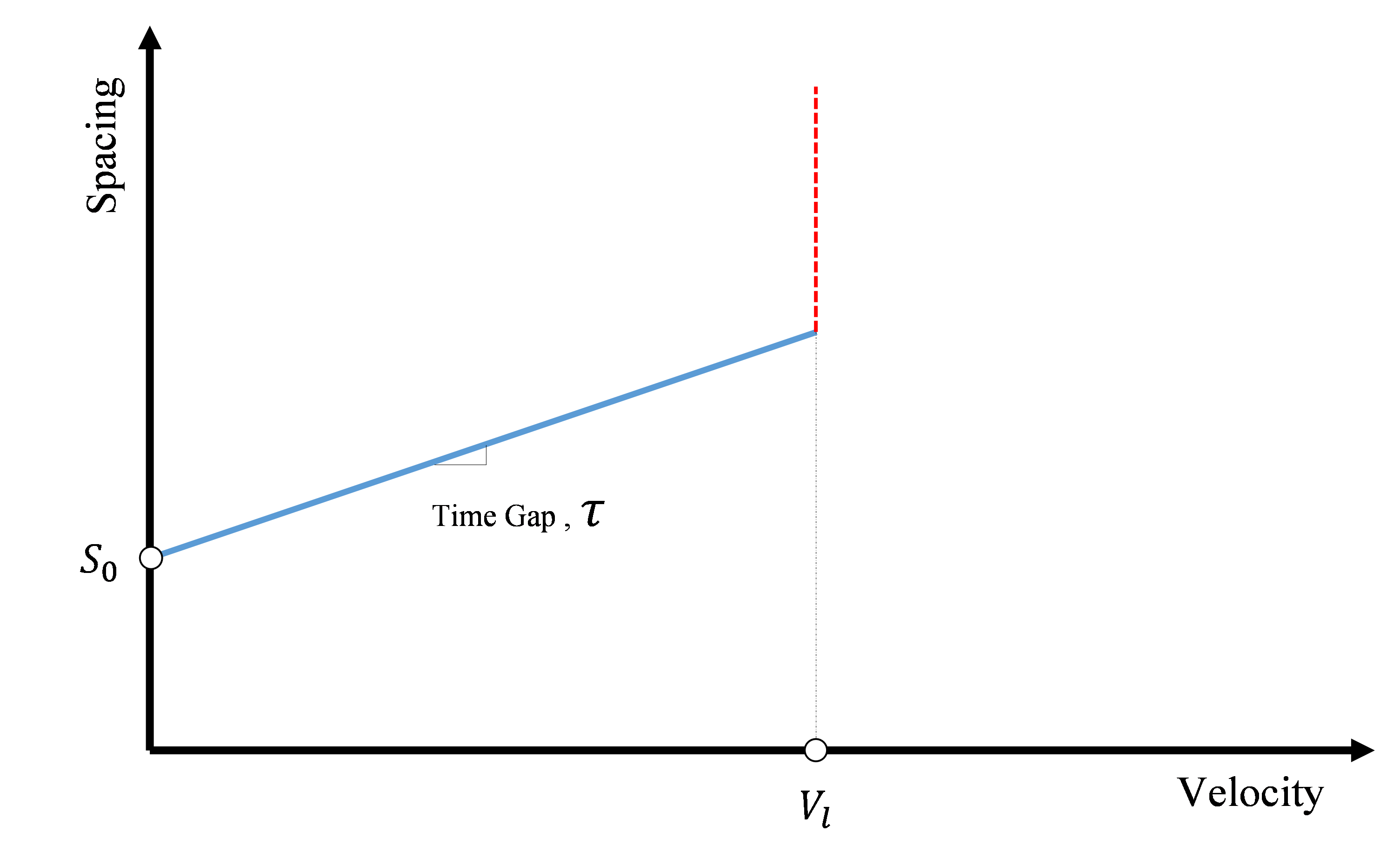}}
    \caption{Relationship between spacing and velocity}
    \label{fig:cth}
\end{figure}

The upper level controller illustrated in Fig. \ref{fig:illustration} commands the car-following behavior of the automated vehicle by regulating the spacing between the follower and its leader. The spacing is based on the CTH policy, which shadows Newell's car-following model in relating the spacing linearly with the speed of the vehicle (see Fig. \ref{fig:cth}). Such formulation depicts the natural driving behavior, where cars slow down when the spacing decreases. Accordingly, a key element of controlling the longitudinal movement of an AV is assigning a spacing that the vehicle should follow. 

Thus, using the CTH policy, the desired spacing is determined as follows: 

\begin{equation}
\label{eq:spacing}
s_i^d(t) = v_i(t)\times \tau^* + s_0
\end{equation}
\begin{equation}
\Delta s_i(t) = s_i(t) - s_i^d(t)
\end{equation}
where $s_i^d(t)$ is the desired spacing of vehicle $i$ at time $t$; $v_i(t)$ is the speed of vehicle $i$ at time $t$; $\tau^*$ is a pre-defined constant time gap; and $s_o$ is the standstill spacing of vehicle i. $\Delta s_i(t)$ is the deviation from the desired spacing of vehicle $i$ at time $t$; and $s_i(t)$ is the actual spacing of vehicle $i$ at time $t$. 

Consequently, the vehicle controller is set to regulate the acceleration rate at every instant $t$ to realize a minimum deviation from the desired spacing. In optimal conditions, where the vehicle can perfectly follow the desired spacing, we expect the actual time gap between consecutive vehicles to be equal to the pre-defined time gap; $\tau^*$. While this is a desired situation, factors such as environmental noise, communication delays and measurement errors will lead to variations between the actual time gap and the pre-defined time gap setting. Variations in time gap could be stemming from these uncertainties, yet communication delays and measurement errors are exogenous, in contrast those related to time gap are endogenous. Previous studies have incorporated exogenous uncertainties in the formulation of lower level controller \cite{wang2018delay,wang2014rolling}, yet explicit uncertainties in time gap variations have not been studied. Notably, decreasing the variations in time gap is specifically of importance in current ACC/CACC control systems, as large variations may lead to driver discomfort, loss of stability (e.g., time gap overshooting along the platoon) and performance hindrance.

\subsection{Model Development}

In this section, we model the spacing between the leading and following vehicles as a random effects model. Specifically, the proposed model follows a parametric approach with random coefficients. This is particularly suitable for our aim to describe the variations in the time-gap parameter while preserving the functional form of the CTH. The random coefficients also allow for describing the vehicle specific variations as well as variations within a platoon of vehicles. 

Without loss of generality, we build upon the functional form of CTH explained above to represent the spacing as follows:

\begin{equation}
    S_i(t) = \mathcal{V}^T(t) \times \Gamma_i + \epsilon_i(t)
\end{equation}
where $S_i(t)$ is the spacing between vehicle $i$ and $i-1$; $\mathcal{V}^T$ contains the intercept and the speed measurements; $\Gamma_i$ is a vector of random coefficients; and $\epsilon_i(t)$ is an error term introduced to capture measurement errors, environmental noise, etc. The error term is assumed to be independent and follows a normal distribution $\mathcal{N}(0,\sigma^2)$. As for the random coefficients $\Gamma_i$, we assume they follow a multi-variate normal distribution $\mathcal{N}(\mu_b,\Sigma_b)$. Previous studies have shown that the model performance is generally robust against misspecifications in the distribution of these random variables \cite{hsieh2006joint,elwany2009real}. By expressing $\mathcal{V}^T(t) = [1,V_i(t)]$ and $\Gamma_i = [s_0, \tau_i]^T$, where $V_i(t)$ represents the speed measurements of vehicle i; $s_0$ is the standstill spacing; and $\tau_i$ is the time gap. The spacing formulation will lead to the following form

\begin{equation}
\label{eq:model}
S_i(t) = \mathcal{V}^T(t) \times \Gamma_i + \epsilon_i(t) = s_0 + \tau_iV_i(t) + \epsilon_i(t)
\end{equation}

\begin{figure}[!htb]
    \centering
    \centerline{\includegraphics[width=0.6\columnwidth]{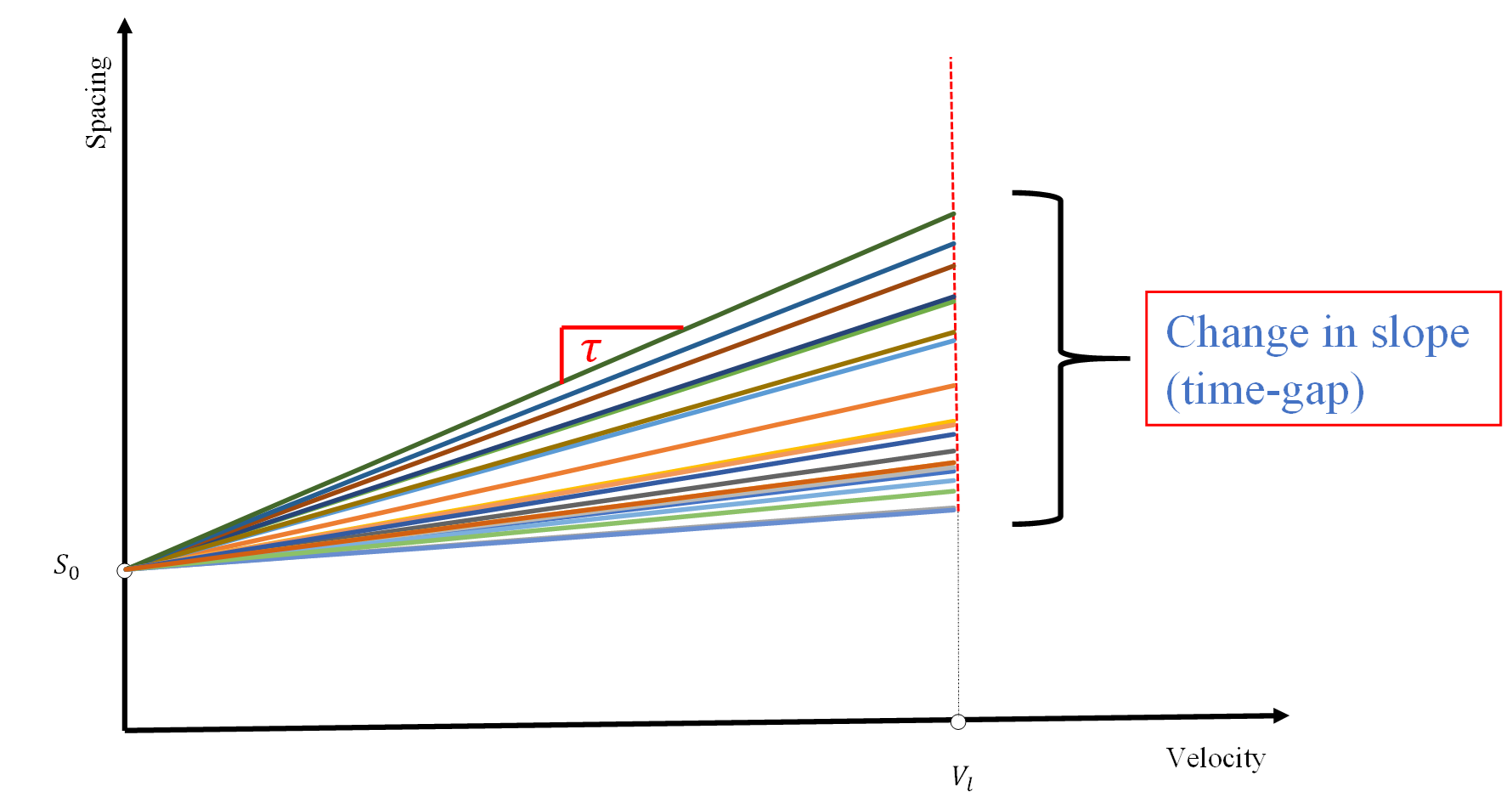}}
    \caption{Spacing formulation for random time gap coefficient}
    \label{fig:random}
\end{figure}

Specifically, by allowing $\Gamma_i$ to be random, we can explicitly account for variations in the time gap parameter, thereby enabling real-time monitoring. An illustration of the time gap as a random coefficient is shown in Fig. \ref{fig:random}.

\subsection{Time Gap Distribution Updating using Bayesian Statistics}

The variations in time gap parameter are attributed to the performance of the vehicle controller in real-time where the vehicle is subject to frequent disturbances and system uncertainty. Thus, to monitor these variations it is essential to estimate the parameters through real-time inference from sensor data. For this, we use a Bayesian updating scheme where both prior knowledge and real time vehicle-specific information are fused together for estimation. Typically, Bayesian updating is a two-stage process: offline and online. In offline stage, a prior knowledge of the parameter is specified either based on historical data or expert knowledge. In our case, we choose the desired time gap as the prior knowledge since the actual time gap is expected to be equal to the desired value. In the online stage, we draw on the advantage of real-time data gathered by vehicle sensors (e.g., LIDAR) to update the estimates of the random coefficients in the model developed above. Details follow

We assume that by instance $t^*$, $n$ measurements of the spacing and speed of vehicle $a$ are gathered from on-board vehicle sensors, denoted respectively as $S^*_a = \{S_{a1}, S_{a2}, S_{a3}, S_{a4}, ..., S_{an} \}^T = \{S_a(t_1), S_a(t_2), ..., S_a(t_n)\}^T$ and $V_a^* = \{V_{a1}, V_{a2}, V_{a3}, V_{a4}, ..., V_{an}\}^T$, where $t_n \leq t^*$. Setting the observed data in Eq. \ref{eq:model} we get: (note: here the asterisk denotes real-time observed values)
\begin{equation}
    S_a^*  = Z_a^* \times \Gamma_a + \mathcal{E}_a^*
\end{equation}
where $\mathcal{E}_a^* = \{\epsilon_a(t_1), \epsilon_a(t_2), ..., \epsilon_a(t_n)\}$ and $Z_a^* = \begin{bmatrix} \mathcal{V}^T_a(t_1)\\...\\\mathcal{V}^T_a(t_n) \end{bmatrix}$

Thus, we use the prior information of $\Gamma_i \sim \mathcal{N}(\mu_b, \Sigma_b)$, where $\mu_b = [\mu_{s_0}, \mu_\tau]^T$ is a $2\times1$ matrix with $\mu_{s_0}$ and $\mu_\tau$ representing the mean values of standstill spacing and actual time gap respectively. $\Sigma_b$ is a $2\times2$ covariance matrix associated with $s_0$ and $\tau$, to compute the posterior distribution $\Gamma_a$ according to the Bayes theorem:

\begin{equation}
     p(\Gamma_a | S^*_a) \propto p(S_a^*|\Gamma_a)\pi(\Gamma_i) 
\end{equation}

where $\pi(\Gamma_i)\sim \mathcal{N}(\mu_b,\Sigma_b)$ represents the prior distribution of $\Gamma_a$. Consequently, we present the likelihood function of $p(S_a^*|\Gamma_a)$ as:

\begin{equation}
\label{eq:likelihood}
    p(S_a^*|\Gamma_a) = \prod^{n}_{j=1}{p(S_{aj}|\Gamma_a)} = \prod^{m}_{j=1}{\frac{1}{\sqrt{2\pi\sigma^2}}\times\ exp\{\frac{-1[S_{aj}-\mathcal{V}_a^T(t_j)\Gamma_a]^2}{2\sigma^2}\}}
\end{equation}

\noindent \textbf{\emph{Proposition}}: The posterior distribution $p(\Gamma_a|S^*_a)$ is a multivariate normal distribution: i.e., $p(\Gamma_a|S_a^*) = \mathcal{N}(\mu_a^*, \Sigma^*_a)$, where: 

\begin{equation}
    \begin{cases}
    \mu^*_a &= \Sigma_a^* \Bigg[ \frac{Z_a^{*^T} S_a^*}{\sigma^2} + \Sigma_b^{-1} \mu_b\Bigg]  \\
    
    \Sigma_a^* &= \Bigg[\Sigma_b^{-1} + \frac{Z_a^{*^T}Z_a^*}{\sigma^2} \Bigg]^{-1}
    \end{cases}
\end{equation}

\noindent \textbf{\emph{Proof:}} The likelihood function in Eq. \ref{eq:likelihood} can be written as: 

\begin{equation}
    p(S_a^*|\Gamma_a) = (2\pi\sigma^2)^{-n/2} \times \text{exp}\big[\frac{-(S^*_a - Z_a^*)^T(S_a^* - Z_a^*\Gamma_a)}{2\sigma^2}\big]
\end{equation}

Based on the prior distribution $\pi(\Gamma_i) = \mathcal{N}(\mu_b, \Sigma_b)$, the posterior distribution of $\Gamma_a$ can the be formulated as: 
\begin{align}
p(\Gamma_a|S_a^*) &\propto p(S_a^*|\Gamma_a)\pi(\Gamma_i)\\
&\propto \text{exp} \Bigg\{\frac{-1}{2\sigma^2} (S_a^* - Z_a^*\Gamma_a)^T(S_a^* - Z_a^*\Gamma_a) \Bigg\} \times \text{exp} \Bigg\{ \frac{-1}{2}\Bigg[(\Gamma_a - \mu_b)^T\Sigma_b^{-1}(\Gamma_a - \mu_b) \Bigg] \Bigg\}\\
&\propto \text{exp} \Bigg\{ \frac{-1}{2\sigma^2} \Bigg[ S_a^{*^T}S_a^* - \Gamma_a^TZ_a^{*^T}S_a^* - S_a^{*^T}Z_a^*\Gamma_a + \Gamma_a^TZ_a^{*^T}Z_a^*\Gamma_a \Bigg]\Bigg\}\\
&\times \text{exp} \Bigg\{\frac{-1}{2} \Bigg[\Gamma_a^T\Sigma_b^{-1}\Gamma_a - \mu_b^T\Sigma_b^{-1}\Gamma_a - \Gamma_a^T\Sigma_b^{-1}\mu_b + \mu_b^T\Sigma_b^{-1}\mu_b \Bigg]\Bigg\}\\
&\propto \text{exp} \Bigg\{\frac{-1}{2}\Bigg[ \Gamma_a^T\Big( \frac{Z_a^{*^T}Z_a^*}{\sigma^2} + \Sigma_b^{-1}\Big)\times\Gamma_a  - \Gamma_a^T\Bigg(\frac{Z_a^{*^T}S_a^*}{\sigma^2} + \Sigma_b^{-1}\mu_b\Bigg) - \Bigg(\frac{S_a^*Z_a^{*^T}}{\sigma^2 + \mu_b^T\Sigma_b^{-1}}\Bigg)\Gamma_a + \mathcal{U}_1\Bigg] \Bigg\}\\
&\propto \text{exp} \Bigg\{ \frac{-1}{2}\Bigg[\Psi^T \Bigg( \frac{Z_a^{*^T}Z_a^*}{\sigma^2} + \Sigma_b^{-1}\Bigg)\Psi +\mathcal{U}_2 \Bigg]      \Bigg\}
\end{align}

Here $\mathcal{U}_1$ and $\mathcal{U}_2$ are constants, and $\Psi$ is a vector defined as: 

\begin{equation}
    \Psi = \Gamma_a - \Bigg(\frac{Z_a^{*^T}Z_a^*}{\sigma^2} + \Sigma_b^{-1}\Bigg)^{-1} \Bigg( \frac{Z_a^{*^T}S_a^*}{\sigma^2} + \Sigma_b^{-1}\mu_b\Bigg)
\end{equation}

Accordingly, we can define the following:

\begin{align}
    \begin{cases}
    \mu_a^* &= \Big( \frac{Z_a^{*^T}Z_a^*}{\sigma^2} + \Sigma_b^{-1}\Big)^{-1} \Big( \frac{Z_a^{*^T}Z_a^*}{\sigma^2} + \Sigma_b^{-1}\mu_b \Big)\\
    \Sigma_a^* &= \Big( \frac{Z_a^{*^T} Z_a^*}{\sigma^2} + \Sigma_b^{-1}\Big)^{-1}
    \end{cases}
\end{align}

Thus, the result can be rewritten as: 

\begin{equation}
\label{eq:density}
    p(\Gamma_a|S_a^*)\propto \Big\{\frac{-1}{2}\big[ (\Gamma_a - \mu_a^*)^T\Sigma_a^{*^{-1}}(\Gamma_a - \mu_a^*)\big]\Big\}
\end{equation}

The probability density function in Eq. \ref{eq:density} is then a multivariate normal distribution $\mathcal{N}(\mu_a^*, \Sigma_a^*)$. This concludes our proof. 

The formulation provided in Eq. \ref{eq:density} presents a closed form Bayesian updating scheme, which reduces drastically the computational load thanks to the desired analytical advantages. This allows our monitoring system to be used on-board the autonomous vehicle. 

\subsection{Monitoring through Control Charts}

We combine the Bayesian updating scheme with Shewart univariate control charts to monitor the variations in the time gap parameter in real-time. This monitoring methodology is particularly beneficial in the ACC/CACC systems, where the user would have the preference in choosing a desired time gap setting (e.g., in \cite{shladover2009effects} users can select their desired time gap setting). Specifically, if large variations in the time gap are observed due to an undesirable event (e.g., leader human driven vehicle is experiencing a speed disturbance), our monitoring methodology can suggest a change in time gap setting to stabilize the variations.

In our framework, an undesirable situation that manifests itself as a significant change in the time gap distribution is detected systematically by the Shewart control chart. In this method, a baseline distribution of the time gap parameter is designed, and then a distance metric is used to measure the deviations from the baseline distribution \cite{hsieh2006joint}. We design the baseline distribution according to some preferred values for the parameter setting. For instance, the baseline distribution of time gap would be a normal distribution, $\mathcal{N}(\mu_{desired}, \sigma_{desired})$, where $\mu_{desired} = \tau^*$ (time gap setting) and $\sigma_{desired}$ is the acceptable variation in time gap. Accordingly, lower and upper bounds are computed to define the acceptable domain of variations:

\begin{align}
\label{eq:limits}
    \mbox{Lower Control Limit (LCL)} &= \mu_{desired} - \mathcal{L}\times\sigma_{desired}\\
    \mbox{Center Line (CL)} &= \mu_{desired}\\
    \mbox{Upper Control Limit (UCL)} &= \mu_{desired} + \mathcal{L}\times\sigma_{desired}
\end{align}

Here $\mathcal{L}$ is a multiplicative value representing the desired confidence level (i.e., within $\sigma$,$2\sigma$, or $3\sigma$). This allows us to compare the updated time gap estimated through the Bayesian framework with the control limits to detect when the time gap goes out of these bounds, triggering a change in the time gap setting.

\section{Application Analysis}
In this section, we demonstrate the application of our monitoring methodology through a simulation experiment, incorporating the real vehicle trajectory data (NGSIM data). Specifically, our designed scenario involves a leading human-driven vehicle followed by a platoon of five CAVs. We monitor the variation of time gap over time for the five CAVs and detect any undesirable events.

\subsection{Scenario Design}
\begin{figure}[!htb]
    \centering
    \centerline{\includegraphics[width=0.8\columnwidth]{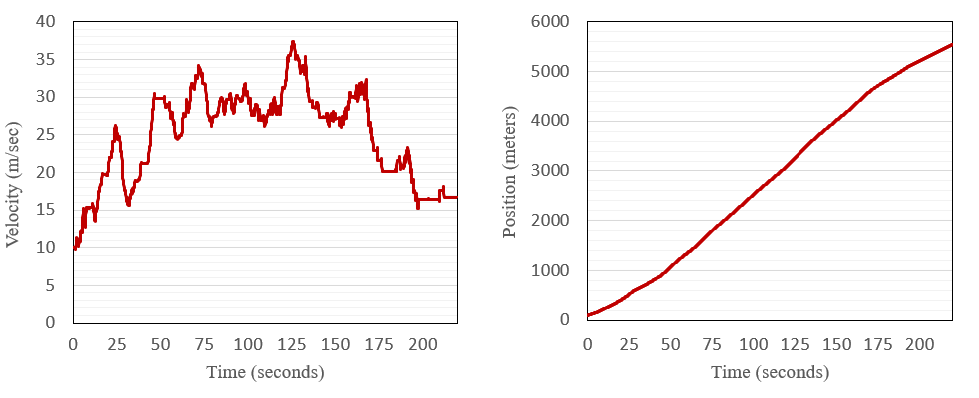}}
    \caption{Scenario design: (left) velocity profile; (right) trajectory}
    \label{fig:scenario}
\end{figure}

Our aim is to study the variations in time gap when an autonomous vehicle (could be with or without communication abilities) is following a human-driven vehicle undergoing aggressive cycles of acceleration/deceleration in the speed range of $20\mbox{mph} - 80\mbox{mph}$. To do so, we extract the acceleration profile of vehicle 1829 from NGSIM dataset for I-80 \cite{ngism} and create a vehicle trajectory for the desired velocity range. The simulated trajectory was created by assuming an initial velocity and location, then using the acceleration profile to construct the trajectory path. This was specifically done in order to study how the monitoring method developed in this paper would perform when the controlled vehicle is subjected to vast disturbances. The variations present in the acceleration/deceleration profile will help mimic uncertainty that could arise in real conditions due to endogenous factors, exogenous factors and uncertain driving behavior. The full scenario design is illustrated in Fig. \ref{fig:scenario}. 

\subsection{Controller Used}
The controller used here is the linear controller developed by Zhou and Ahn \cite{zhou2019robust}. Here we briefly summarize the controller, and readers are referred to the cited paper for more details. The controller’s upper level follows the CTH policy to compute the desired spacing as in Eq. \ref{eq:spacing}, and the lower level controller incorporates the generalized vehicle dynamics (GLVD):

\begin{equation}
    \dot{a}(t) = \frac{-1}{T_{i}} a_i(t) + \frac{K_i}{T_i}u_i(t)
\end{equation}
where $\dot{a}(t)$ i the jerk; $a_i(t)$ is the actual acceleration rate realized for vehicle i; $u_i(t)$ is the acceleration rate commanded by the upper level controller; $K_i$ is the ratio of the commanded acceleration to the realized acceleration for vehicle $i$; and $T_i$ is the actuation time lag. Accordingly, the system can be formulated as a state-space as follows: 

\begin{equation}
    \dot{x}_i(t) = A_ix_i(t) + B_iu_i(t) + Da_{i-1}(t)
\end{equation}

where $x_i(t) = [\Delta d_i(t), \Delta v_i(t), a_i(t)]^T$; $\Delta d_i(t)$ is the deviation from desired spacing; $\Delta v_i(t)$ is speed difference between vehicle $i$ and $i-1$; $A_i = \begin{bmatrix} 0&1&\tau^*\\0&0&-1\\0&0&\frac{-1}{T_i}\end{bmatrix}$; $B_i = \begin{bmatrix}0\\0\\K_i/T_i\end{bmatrix}$; $D_i = \begin{bmatrix}0\\1\\0\end{bmatrix}$

Then, the controller demanded acceleration is:

\begin{equation}
u_i(t) = k_ix_i(t) + k_{fi}a_{i-1}(t-\theta)
\end{equation}
here $k_i$ is a vector of control gains; $k_{fi}$ is a coefficient for feedforward gain; $\theta$ is a variable for communication delay. 

The parameter values used in the simulation experiment are shown in Table \ref{tab:setup}. We chose the desired time gap of 1.6 seconds as it is one of the available setting options in the current ACC/CACC systems. Based on this parameter setting, we simulate CAV trajectories using the above controller.  

\begin{table}[!ht]
\centering
\caption{Default parameter settings for simulation setup}
    \begin{tabular}{ll}
    \hline
    \textbf{Parameters} & \textbf{Value}\\
    \hline
        $T_i$ & 0.45 secs\\
        $K_i$ & 1 \\ 
        Run time & 250 secs \\
        Time step & 0.1 sec\\
        $\tau^*$ & 1.6 sec \\
        $s_0$ & 5 m\\
    \hline
\end{tabular}
\label{tab:setup}
\end{table}

\subsection{Monitoring Profiles and Control Charts}

Our monitoring methodology is then coded into an algorithm allowing to profile the time gap of every vehicle over time. Also, we determine the acceptable domain of variations through the Shewart control charts. We compute the control limits based on a $95\%$ confidence level (i.e., $2 \sigma $). We assume that our baseline time gap distribution is $\mathcal{N}(1.6, 0.125)$. This means that we expect our actual time gap to be $1.6$ while accepting a standard deviation of $0.125$. Then we can compute the control limits as in Eqs. 19-21. Thus, we obtain: $\mbox{LCL} = 1.35$, $\mbox{CL} = 1.6$, and $\mbox{UCL} = 1.85$.

Without loss of genrality, we assume $\epsilon=\mathcal{N}(0,0.01)$ and the prior distribution of random coefficients in the Bayesian updating as $\Gamma_i = \mathcal{N}(\mu_b, \Sigma_b)$, where:

\begin{align}
    \begin{cases}
    \mu_b &= [1, 1.6]^T\\
    \Sigma_b &= \begin{bmatrix} 0.0001 & -1e-5\\ -1e-5 & 0.125
    \end{bmatrix}
    \end{cases}
\end{align}

\begin{figure}[!htb]
    \centering
    \centerline{\includegraphics[width=1\columnwidth]{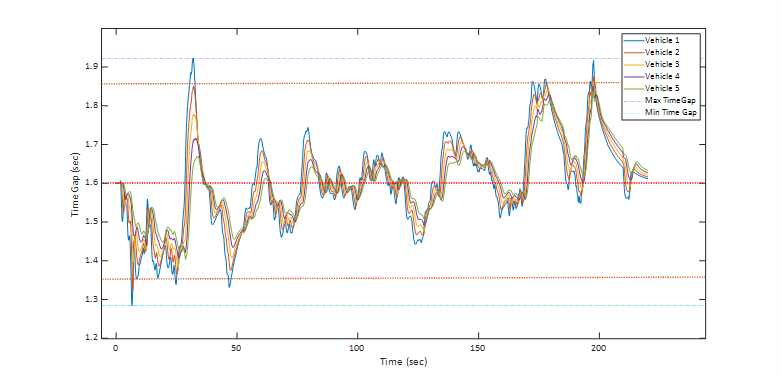}}
    \caption{Time gap profiles and control limits for vehicles 1 to 5}
    \label{fig:1}
\end{figure}

Fig. \ref{fig:1} presents the time gap profiles of the vehicles over time along with the control limits (red lines). The analysis shows that the variations exceed the limits for vehicles 1 and 2, where the max time gap reaches 1.92 seconds and minimum of 1.28 seconds. Furthermore, the time gap profile shows significant variations within a short time (going out of the bounds four times within 50 seconds), suggesting undesirable performance. Vehicles 3, 4, and 5 show a decrease in variations, which is expected as the controller is designed to dampen disturbances along the platoon to ensure string stability. 

A significant advantage by our real-time monitoring methodology is to support real-time parameter adjustment to improve performance. For instance, once we detect large variations in the time gap, we can change the time gap setting to realize lower variations. For our example, after we detect that the time gap of vehicle 1 exceeded the control limits three times within 35 seconds, we change the desired time gap from 1.6 seconds to 1 second. This decrease in the time gap helps in dampening the disturbances more effectively. Now, we change our control limits to adjust to the new time gap setting: $\mbox{LCL} = 0.75$, $\mbox{CL} = 1$, and $\mbox{UCL} = 1.25$ (based on Eqs. 19-21)

\begin{figure}[!htb]
    \centering
    \centerline{\includegraphics[width=1\columnwidth]{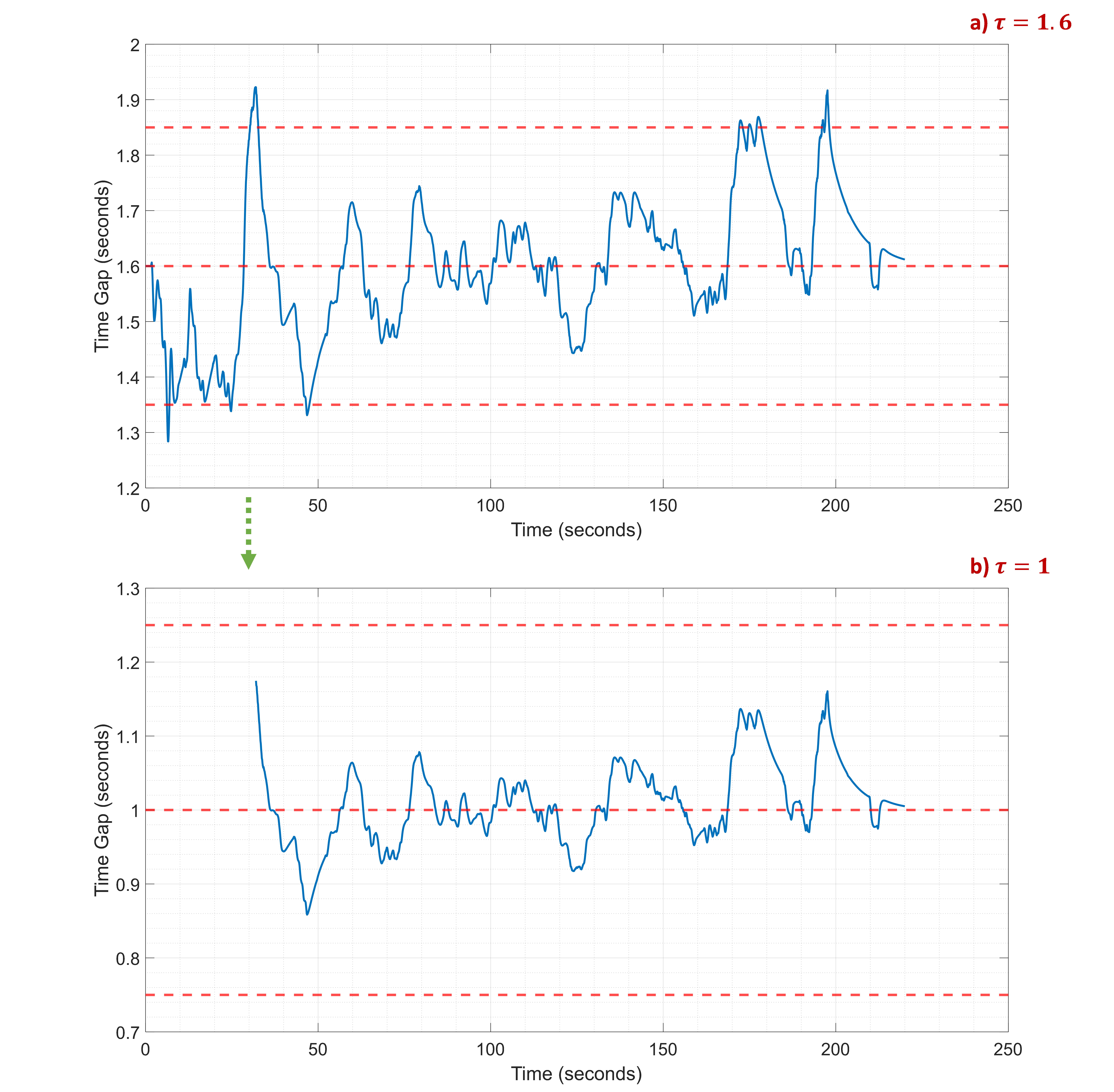}}
    \caption{Vehicle 1 time gap profile before and after change; (upper) time gap 1.6; (lower) time gap 1}
    \label{fig:2}
\end{figure}

Fig. \ref{fig:2} demonstrates that the change in the time gap setting leads to drastically lower variations in time gap for vehicle 1: where the maximum deviation from the desired time gap is reduced to 0.17 from 0.32 when the time gap was set at 1.6 seconds. Additionally, upon inspecting the entire platoon profile before and after the change. An interesting observation is that for vehicles 3, 4 \& 5 the variations under both time gaps are small. Thus, it is viable to only change the time gap setting for vehicles experiencing large variations.

\section{Conclusions}

This study presented a novel real-time monitoring methodology for time gap variations informed from vehicle sensor data on spacing and speed. The main contributions of this study are: (1) motivating the importance of monitoring time gap variations as a key performance metric for vehicle’s control system; (2) developing a formulation of the spacing between vehicles that addresses the stochastic nature of time gap parameter through incorporating random coefficients; (3) providing derivation and proof of a closed-form Bayesian updating scheme that reduces the computational load and enables real-time implementation; (4) incorporating control charts in the monitoring scheme to alert when a change in time gap is desired. 

Furthermore, the study showcased application of the monitoring methodology through simulations utilizing the NGSIM data. We monitor the time gap profile of a platoon of CAVs following a human driven vehicle undergoing cycles of aggressive deceleration and acceleration, and the results showed that the variation in time gap exceeded the desired limits. When the time gap setting was changed as informed by our monitoring system, the variation decreased significantly, demonstrating the effectiveness of the proposed monitoring system.  

Nevertheless, this study can be enhanced in several ways. Real experimental data on autonomous vehicles can be used to systematically analyze the uncertainty in time gap. A non–linear modeling approach can also be considered to obtain more accurate estimates of time gap in real-time. Furthermore, the time gap parameter depends on key control parameters such as feedback and feedforward gains, which are not considered in this work. Finally, incorporation of other performance metrics will result in a better monitoring methodology.

\section*{Acknowledgments}
This research was sponsored by the United States National Science Foundation through Award CMMI 1536599 and the University of Wisconsin Madison

\bibliographystyle{unsrt}  
\bibliography{references}  
\end{document}